# The Normal State Resistivity of Grain Boundaries in $YBa_2Cu_3O_{7-\delta}$

J. H. T. Ransley,[1] S. H. Mennema,[1] K. G. Sandeman,[1] G. Burnell,[1] J. I. Kye,[2] B. Oh,[2] E. J. Tarte,[1] J. E. Evetts[1] and M. G. Blamire[1]

[1] IRC in Superconductivity, Madingley Road, Cambridge, CB3 0HE, UK

[2] LG Electronics institute of Technology, Seoul, 137-724, Korea

Using an optimized bridge geometry we have been able to make accurate measurements of the properties of $YBa_2Cu_3O_{7-\delta}$ grain boundaries above $T_c$. The results show a strong dependence of the change of resistance with temperature on grain boundary angle. Analysis of our results in the context of band-bending at the boundary allows us to estimate the height of the potential barrier present at the grain boundary interface.

The discovery of high critical temperature ($T_c$) superconductivity[1] was followed rapidly by the realization that the superconducting currents attainable in practical, polycrystalline materials are severely reduced by grain boundaries (GBs).[2,3] The nature of these GBs is a subject of considerable interest[3-5] but, in spite of the importance of these defects, a consensus as to their true nature has not emerged and a number of conflicting models exist. Although the GB microstructure is now well characterized,[6,7] the relationship between microstructure and electronic structure remains a matter of dispute. For example, the observation of electromagnetic resonances in the boundary[8] has confirmed that it is, at least in part, electrically insulating. Some authors argue that these insulating regions are a result of a metal-insulator transition, caused by disruption of the carrier transfer to the copper oxide planes[9] or severe oxygen depletion[10] (in this letter we will refer to such a scenario as 'intrinsically insulating'). Others argue that band-bending,[11] a local distortion of the band structure on a smaller energy scale, occurs. In a band-bending picture the GB is charged, and produces an electric field that significantly shifts the potential of the local band structure; this has been observed at GBs in the structurally similar oxide $SrTiO_3$.[12] If the extent of the band-bending is such that the bottom or top of the band passes through the Fermi level then the material becomes electrically insulating in the region of the boundary and the electrons must tunnel through the barrier. In such an event, the shape[11] and height of the tunnel barrier would be significantly different from that characteristic of traditional metal-insulator-metal barriers and, critically, the barrier height will be much less than would be the case if the material were 'intrinsically insulating'. This is because the barrier height in a band-bending scenario is determined by the charge on the boundary, whilst in other cases it is given by the energy difference between the top of the band relevant for transport and the next empty band, which is of order 1 eV.[13] Here we present results from a measurement technique which enables the determination of the height and shape of the electrical barrier at the GB; the results are consistent with a band-bending model and may help clarify the true nature of the GB barrier.

Normal-state measurements of GBs have not been performed until now because of the relatively large resistance of the tracks; the resistance of a $0.1 \mu m \times 2 \mu m$ $YBa_2Cu_3O_{7-\delta}$ (YBCO) 24° [100] tilt misoriented boundary (approximately 4 Ω at 290 K) is significantly less than the resistance of a reasonable length of adjoining track (33 Ω for a 0.1 $\mu$m thick, square section at 290 K). We used a Wheatstone bridge structure[14] to measure the temperature dependence of the grain boundary resistance (shown in Fig. 1); the symmetry of the structure ensures that all the resistance contributions balance to zero except for those arising from the GB.

In practice any fabricated bridge structure contains imperfections, so the measured imbalance signal always has some component due to asymmetries in the structure. As Fig. 1(a) shows, the lithographic process used to fabricate the bridges (in this case photolithography and Ar ion milling) produces inhomogeneities on the micrometer length scale. On a larger scale, it is difficult to deposit thin films that have perfectly uniform properties over hundreds of micrometers (in this study the films were grown by pulsed laser deposition on $SrTiO_3$ bicrystal substrates). The imbalance due to these two sources of error can be estimated by fabricating bridges from films without grain boundaries; the signal obtained from such bridges can be analyzed in the same way as the imbalance due to a grain boundary, to produce an equivalent resistance area product and so determine the error on the GB resistance measurement. By experimenting with different bridge geometries we have found that the structure shown in Fig. 1(b) produces a good compromise between these two sources of error.[15]

Using the pattern shown in Fig 1(b), eight such control bridges were fabricated from a epitaxial YBCO film. The mean equivalent resistance area product ($RA$) of these bridges at 250 K was found to be $(-5 \pm 4) \times 10^{-14}$ Ω$m^2$ (measured with a current bias of 5 $\mu$A). This value is small compared with values for a typical GB as shown in Fig. 2. The random errors associated with lithographic imperfections are further reduced by taking the mean of data from a number of GB bridges on the same substrate. Further systematic sources of error due to the finite grain boundary thickness and grain boundary grooving contribute errors of order $5 \times 10^{-14}$ Ω$m^2$ at 300 K. These additional sources of error are comparable to or less than the errors associated with the imperfection in the bridge.[15]

Our methods have allowed us to perform the first systematic investigation of the zero voltage normal state resistance of YBCO grain boundaries. These measurements, together with the theoretical framework which we have developed, provide the first direct measurement of the barrier height at the interface. Figure 2 shows the zero bias $RA$ vs temperature ($T$) for the grain boundaries investigated in this work. These data represent the mean properties of a number of boundaries (42 in total per bridge, and typically 5 bridges per sample). While this helps to eliminate systematic errors in the measurement, it should be noted that the spread in the properties of these boundaries is significant; critical current measurements on bridges in which one arm of the device is severed indicate a five fold spread in the critical currents of the boundaries in a single bridge.

As the grain boundary misorientation angle increases, the temperature dependence of the normal state resistance in alters dramatically (see Fig. 2). The strength of the variation with grain boundary angle and temperature is inconsistent with a simple thermal emission process, and with variable range hopping transport. Instead it is more characteristic of a change in the shape of a low-energy tunnel barrier as a result of increased charging and band-bending as the misorientation increases.

The band-bending model would predict that as the carrier density is reduced the depletion length, over which the band structure distorts, extends further into the surrounding material. With this in mind we de-oxygenated the 30° sample by annealing for 7 hours at 500°C in 0.002 atm. of $O_2$ and quenching into liquid nitrogen, reducing its $T_c$ from 90 K to 41 K. Such a treatment is expected to reduce the number of added holes by a factor of approximately two.[16] In Fig. 2 we show that the temperature dependence of the zero bias resistance changes from the characteristic, flat, low angle behavior to the more rapidly varying high angle behavior. Above 100 K a strong similarity with the results from the 36° and 45° samples is apparent and provides further evidence for band-bending. Below this temperature strongly activated behavior is observed, possibly indicative of the opening up of a pseudogap in the density of states (DoS) near the Fermi level.[16,17]

In addition to the normal state properties of the boundaries, we measured their current density-voltage (*J-V*!) characteristics in the superconducting state (measurements of *J-V*!curves in the normal state are seriously affected by heating). The *J-V* data as a function of temperature for the 36° and 45° boundaries are shown in Fig. 3. – the lower angle boundaries had linear *J-V* curves over the accessible voltage range. The *J-V* characteristic in the superconducting state is essentially independent of temperature indicating that heating is not a significant problem; we have chosen to model all the data at 80K, but the shape of both the modeled and experimental data does not depend significantly on temperature below $T_c$.

We have modeled the electron tunneling through the grain boundary potential barrier; full details will be provided in a future publication. Briefly, by considering a conventional tunneling equation[18] for the forward and reverse current we calculate the net current perpendicular to the grain **J** (**J**$_l$–**J**$_r$) from

$$\mathbf{J_l} = \frac{2e}{(2\pi)^3} \int v_F f(\varepsilon)[1-f(\varepsilon+eV)]D_l(\mathbf{k})d\mathbf{k} \quad (1)$$

$$\mathbf{J_r} = \frac{2e}{(2\pi)^3} \int v_F f(\varepsilon+eV)[1-f(\varepsilon)]D_r(\mathbf{k})d\mathbf{k}. \quad (2)$$

where $v_F$ is the Fermi velocity perpendicular to the GB, $\varepsilon(\mathbf{k})$ is the band structure of the metallic region, and $f(E)$ is the usual Fermi occupation function. $D(\mathbf{k})$ is a transmission probability given by the WKB approximation:

$$D(\mathbf{k}) \approx \exp\left(-2\sqrt{\frac{2m}{\hbar^2}} \int_0^w \left(V(x) - \varepsilon(k) + \frac{p_\perp^2}{2m}\right)^{1/2} dx\right) \quad (3)$$

where $V(x)$ is the variation of barrier potential in real space, $p_\perp$ is the momentum perpendicular to the barrier on each side and $m$ is the effective electron mass. $V(x)$ is finite for the width of the barrier, $w$. We choose the following band structure to reflect the in-plane features of the YBCO Fermi surface;

$$\varepsilon(\mathbf{k}) = 2t\left(\cos(k_x a) + \cos(k_y a)\right) + 4t'\cos(k_x a)\cos(k_y a) \\ + 2t''\left(\cos(2k_x a) + \cos(2k_y a)\right) \quad (4)$$

which arises from considering hopping terms up to next nearest neighbor on a square lattice of side $a$; we have used a unit cell with $a = 3.86$ Å and $c = 11.68$ Å

The model has only two free parameters: the height of the boundary potential barrier and the width of the adjacent region over which the band-bending extends (see inset to Fig. 2). Since we have used a relevant band structure for the material,[19] we can predict absolute values of the transport current.

Here we use: $t = -0.52$ eV, $t' = -0.45t$ and $t'' = 0.35t$ with a Fermi level of $-0.693$eV. We assume a trapezoidal barrier profile, with a central section of barrier height $V_h$ of height 0.19 eV above $E_F$ and 'structural width' equal to that of one unit cell (3.86Å) (typical of the range of structural disorder observed in micrographs). With constant barrier height and by varying the width of the total barrier over a small range of values (see inset to Fig. 2) we can fit the *RA* vs *T* curves for all the misorientation angles (Fig. 2). Using the same parameters we can also calculate the *J-V* data; the form of the experimental curves is reproduced. We can also fit the (*J-V!*) data; we have tried to use the highest temperature data to minimize the influence of the superconductivity on the DoS, but in fact the experimental data is almost temperature independent.

The most significant result of this experiment is that the barrier height we determine is much lower than appropriate values for 'intrinsic insulators'[13] and goes a considerable way to demonstrating that band-bending is a plausible mechanism for current suppression at grain boundaries in YBCO. Furthermore, the band-bending model predicts that the boundary width is dependent on the carrier density, a behavior confirmed by our deoxygenation experiment. Because tunneling strongly favors electrons with momentum parallel to the boundary normal it may be possible to investigate the DoS as a function of position on the Fermi surface, using a series of carefully chosen bicrystal geometries and facet-free grain boundaries.[20] The evidence presented here that band-bending is a likely mechanism for current suppression in higher angle boundaries is a significant step forward in our understanding of these technologically important defects.

The authors thank: C. Schneider, J. Adkins, J. Cooper, J. Halbritter, H. Hilgenkamp, P. Littlewood and J. Loram. The work was supported by EPSRC.

*Figure captions*

FIG. 1: The bridge structure used to investigate the normal state properties of the grain boundaries. a) atomic force microscope image of part of a fabricated Wheatstone bridge device. b) Diagram of the entire device structure, showing the line along which the grain boundary is aligned.

FIG. 2: Resistance area product vs. temperature for grain boundaries of different misorientation angles. The data is obtained by averaging results from several bridge structures (illustrated in Fig. 1) on the same substrate. The continuous lines show the experimental data; the dotted lines are calculated from our theoretical model. The grain boundaries measured are 24°(film $T_c$ = 89 K), 30°(film $T_c$ = 92 K) and 36°(film $T_c$ = 82K) symmetric and a 45° asymmetric GB (film $T_c$ = 89 K). Also shown is data for a 30° symmetric GB after de-oxygenation (under-doping) ($T_c$ = 41 K). Insert shows the potential barriers used for the tunneling model.

FIG. 3: Current density – voltage characteristic for single 36° symmetric and 45° asymmetric [001] tilt boundaries. The dashed lines show fits derived from the tunneling model.

Ransley et al. Fig 1

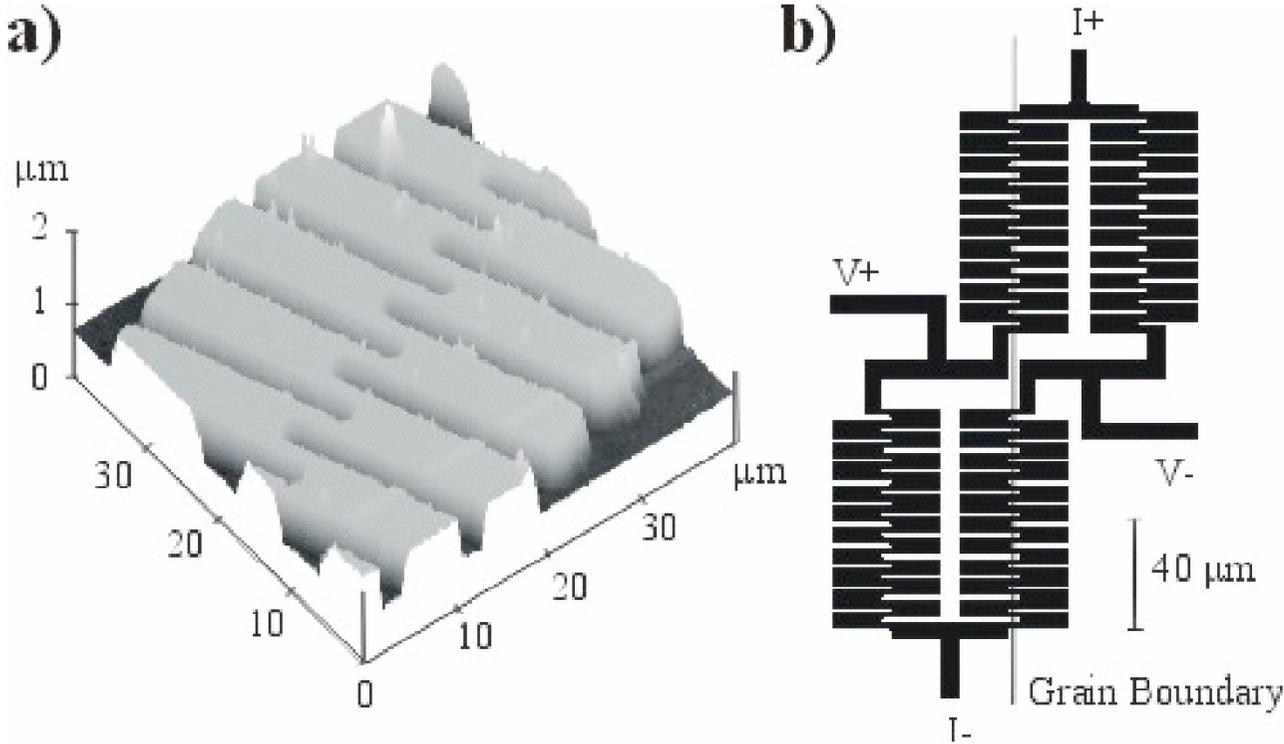

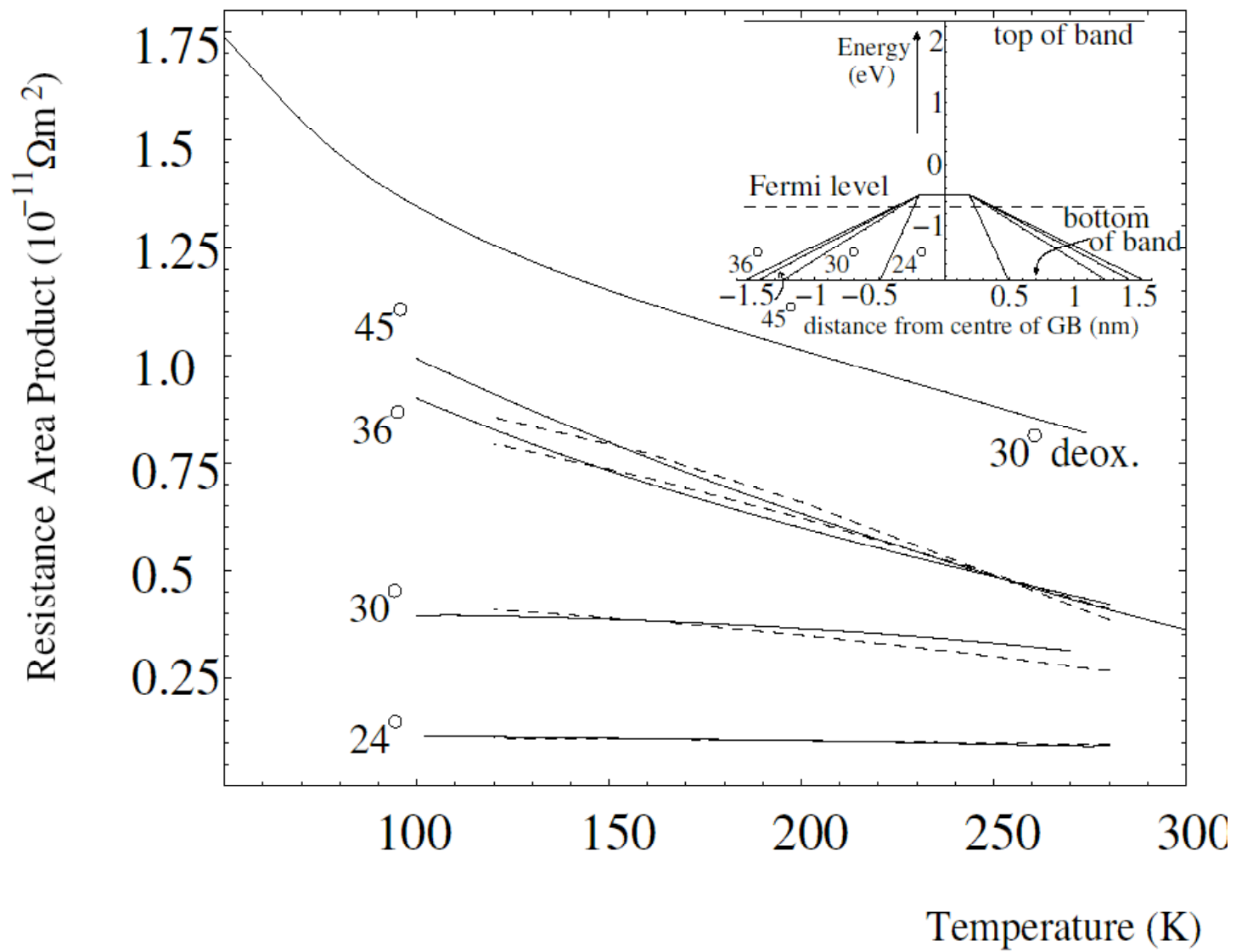

Ransley et al. Fig 3Ransley et al. Fig 3

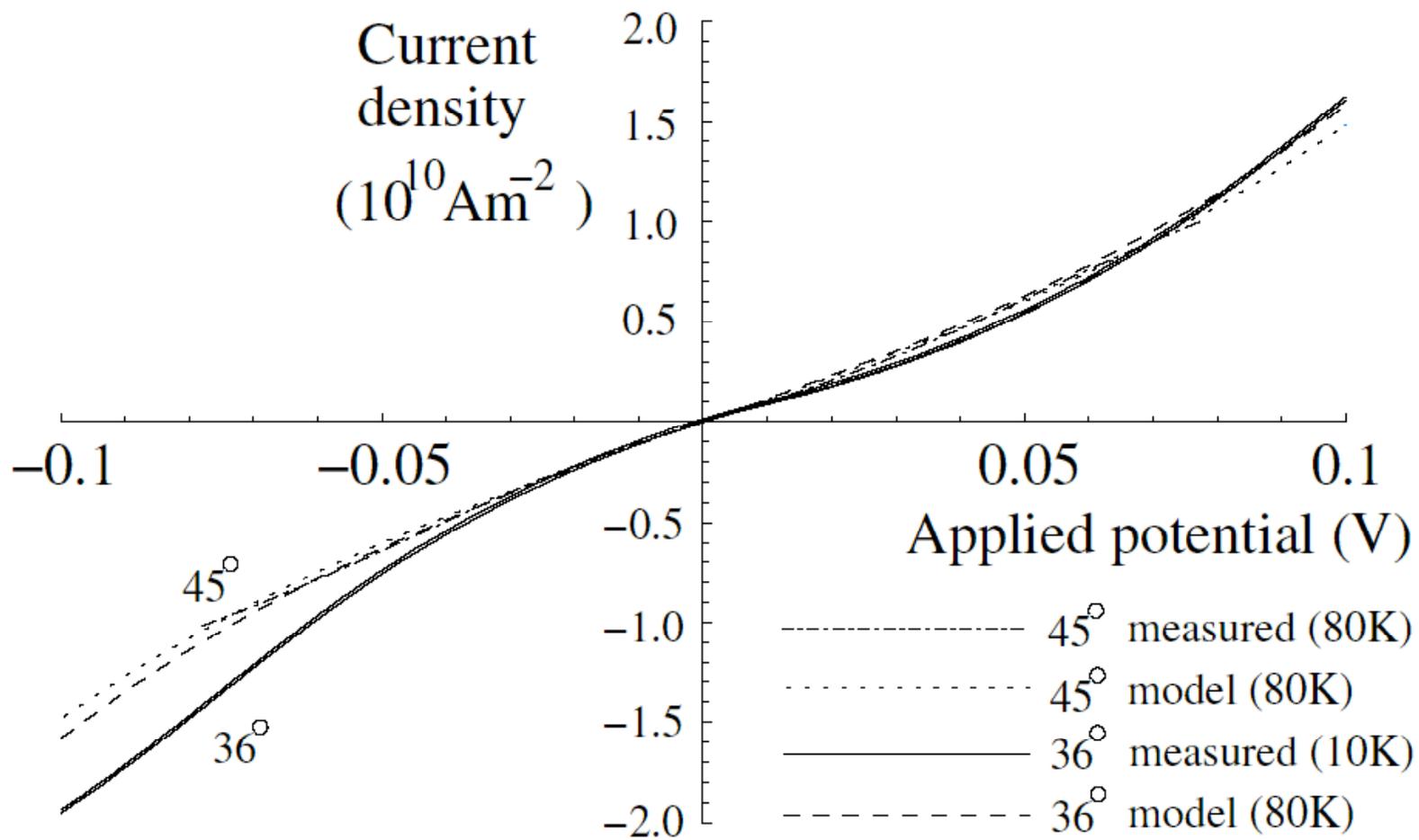